\documentclass[aps,prl,reprint,superscriptaddress]{revtex4-1}
\usepackage{graphicx}% Include figure files
\usepackage{dcolumn}% Align table columns on decimal point
\usepackage{bm}% bold math

\begin{document}

%\preprint{}

\title{Tunable Circularly Polarized Terahertz Radiation from Magnetized Gas Plasma}

\author{W.-M. Wang}
\email{weiminwang1@126.com}\affiliation{Forschungszentrum J\"ulich
GmbH, Institute for Advanced Simulation, J\"ulich Supercomputing
Centre, D-52425 J\"ulich, Germany} \affiliation{Beijing National
Laboratory for Condensed Matter Physics, Institute of Physics, CAS,
Beijing 100190, China}
\author{P. Gibbon}
\affiliation{Forschungszentrum J\"ulich GmbH, Institute for Advanced
Simulation, J\"ulich Supercomputing Centre, D-52425 J\"ulich,
Germany}\affiliation{Centre for Mathematical Plasma Astrophysics,
Katholieke Universiteit Leuven, 3000 Leuven, Belgium}
\author{Z.-M. Sheng}
\affiliation{SUPA, Department of Physics, University of Strathclyde,
Glasgow G4 0NG, United Kingdom}\affiliation{Key Laboratory for Laser
Plasmas (MoE) and Department of Physics and Astronomy, Shanghai Jiao
Tong University, Shanghai 200240, China} \affiliation{IFSA
Collaborative Innovation Center, Shanghai Jiao Tong University,
Shanghai 200240, China}
\author{Y.-T. Li}
\affiliation{Beijing National Laboratory for Condensed Matter
Physics, Institute of Physics, CAS, Beijing 100190, China}
\affiliation{IFSA Collaborative Innovation Center, Shanghai Jiao
Tong University, Shanghai 200240, China}

\date{\today}

\begin{abstract}
It is shown, by simulation and theory, that circularly or
elliptically polarized terahertz radiation can be generated when a
static magnetic (B) field is imposed on a gas target along the
propagation direction of a two-color laser driver. The radiation
frequency is determined by $\sqrt{\omega_p^2+{\omega_c^2}/{4}} +
{\omega_c}/{2}$, where $\omega_p$ is the plasma frequency and
$\omega_c$ is the electron cyclotron frequency. With the increase of
the B field, the radiation changes from a single-cycle broadband
waveform to a continuous narrow-band emission. In high-B-field
cases, the radiation strength is proportional to
$\omega_p^2/\omega_c$. The B field provides a tunability in the
radiation frequency, spectrum width, and field strength.
\end{abstract}


\maketitle

Terahertz (THz) spectroscopy and coherent control have been widely
applied in physics
\cite{THz-spectroscopy,THz-phy1,THz-phy2,THz-phy3}, biology
\cite{THz-bio} and medicine \cite{THz-medicine}. These applications
can potentially benefit from THz radiation sources from gas
\cite{Cook,Kress,Bartel,Xie,Kim,THz_OE,PengXY,Wang_TJ,YChen,Kim2,THz_PRE,THz-COL,THz-PRA}
or solid \cite{YTLi,Gopal2} plasmas irradiated by fs intense laser
pulses thanks to their high radiation strength and bandwidth up to
100 THz. Recently, powerful THz radiation of multi-MV/cm
\cite{WL_Scaling,THz_8MV} has been efficiently generated via a
two-color laser scheme in which a fundamental pump laser is mixed
with its second harmonic in gases \cite{Cook}. Basically, such
radiation generated by linearly-polarized laser drivers is linearly
polarized although, in some conditions, the linear polarization
becomes elliptical during propagation due to modulation of the laser
phase and polarization in gas plasma \cite{Elliptically_polarized}.
To achieve radiation with controllable polarization to further
broaden the THz application scope, e.g., polarization-dependent THz
spectroscopy
\cite{Polarization_app1,Polarization_app2,Polarization_app3},
elliptically polarized (EP) or circularly polarized (CP) laser
pulses have been used to generate EP broadband THz radiations
\cite{WuHC,Dai_CP,Wen_CP}.

In this Letter, we propose a scheme in which a static B field is
imposed along the propagation direction of a two-color linearly
polarized laser driver to generate narrow-band THz radiation of
circular or elliptical polarization with the relative phase between
the two radiation field components fixed at $\pi/2$. The radiation
rotation direction can be controlled by the B-field sign. At a field
strength of 100T, the electron cyclotron frequency
$\omega_c=eB_0/m_ec$ is much higher than the plasma oscillation
frequency $\omega_p=\sqrt{4\pi e^2 n_e/m_e}$ ($n_e$ is the formed
plasma density), so the radiation frequency is almost at $\omega_c$,
and therefore, it can be smoothly tuned by the B-field strength
$B_0$. In this case, the B field dominates over the plasma
oscillation and the former governs the trajectory of the plasma
electrons, which causes such radiation properties. Because of
$\omega_c\gg\omega_p$, the radiation has a many-cycle waveform
rather than a single-cycle waveform
\cite{Cook,Bartel,Xie,Kim,THz_OE,WuHC} in the case without the B
field. Thus, the current radiation has a narrow-band spectrum.

Magnetic fields at tens of teslas are widely available in the form
of dc or ms-pulsed, nondestructive magnets \cite{Labs_highB}, where
the highest one reaches 100T. Via destructive methods, 600T
$\mu$s-pulsed B-fields were available more than a decade ago
\cite{B600T,B310T}. Nanosecond-laser-driven capacitor-coil
experiments demonstrated ns-pulsed B fields of 1500T recently
\cite{Fujioka}, which also has significant applications in novel
magnetically assisted inertial confinement fusion
\cite{MA-FI-PRL,MA-ICF}.

We first demonstrate the scheme sketched above through
particle-in-cell (PIC) simulations with the two-dimensional (2D)
KLAPS code \cite{KLAPS}, in which the field ionization of gases is
included. The pump laser wavelength is fixed at 1$\rm\mu m$ (or the
period $\tau_0=3.33~\rm fs$) and the second laser frequency is at
the second harmonic of the pump one. The two pulses propagate along
the +x direction with linear polarization along the z direction.
They have the same spot radius $r_0=150\rm\mu m$ and duration 50fs
at full width at half maximum. Peak intensity of the pump pulse is
$2\times10^{15}~ \rm W/cm^2$ with the energy 42mJ and the second
pulse has the peak intensity $5\times10^{14}~ \rm W/cm^2$ and energy
11mJ. A helium gas slab is taken with a uniform density $1.22\times
10^{16}~\rm cm^{-3}$ (the corresponding plasma frequency $\omega_p=$
1THz after the complete first-order ionization by the used laser
pulses \cite{THz_PRE}) and a length 320$\rm\mu m$. The resolutions
along the x and y directions are 0.01$\rm\mu m$ and 0.25$\rm\mu m$,
respectively. Initially, in the gas region, four simulation
particles per cell denoting gas atoms are adopted.

\begin{figure}[htbp]
\includegraphics[width=3.2in,height=3.0in]{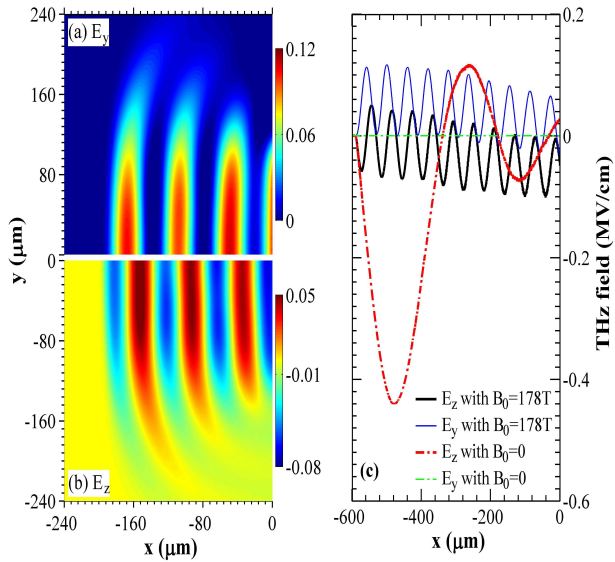}
\caption{\label{fig:epsart}[(a) and (b)] Snapshots of the THz
electric fields (MV/cm) at the time of 0.7 ps and (c) the field
distributions on the axis ($y=0$) at 2 ps, where an external B field
of 178T is imposed. The broken lines in (c) correspond to the case
without the B field.}
\end{figure}

Figure 1 shows spatial distributions of the THz radiation
propagating along the -x direction in the vacuum, which is generated
with an external static B field of 178T imposed along the +x
direction. As a comparison, the radiation generated without the B
field is also displayed by the broken lines in Fig. 1(c),
illustrating that the radiation has only the z-direction component
and a near single-cycle waveform, as shown in previous experiments
and simulations \cite{Cook,Bartel,Xie,Kim,THz_OE}. With the B field,
the radiation also has the y-direction component in addition to the
z-direction one. The two components have the same frequency, higher
than that in the case without the B field, and a constant phase
displacement.

\begin{figure}[htbp]
\includegraphics[width=3.2in,height=2.8in]{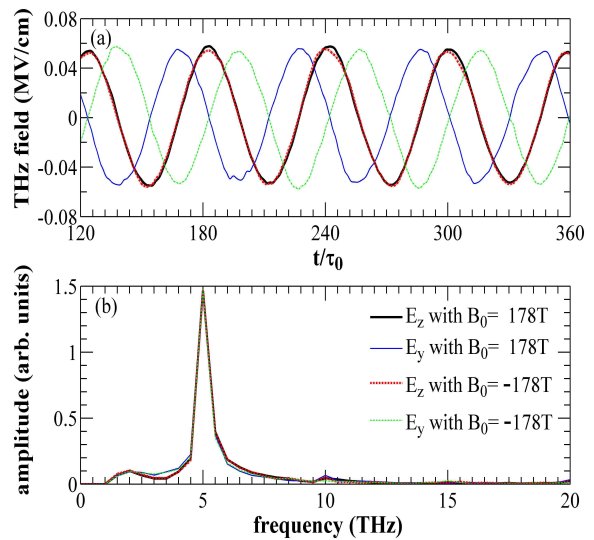}
\caption{\label{fig:epsart}(a) Temporal waveforms of the THz
electric fields on the axis and (b) the corresponding spectra, where
an external B field of 178T is imposed. The components below 1 THz
are filtered.}
\end{figure}

To further analyze the radiation properties, we take the temporal
waveform observed in the left vacuum $5~\rm\mu m$ away from the
vacuum-gas boundary, as illustrated in Fig. 2. The low-frequency
part below 1 THz has been filtered and the high-frequency part is
retained. One can see the radiation frequency at 5 THz, equal to the
cyclotron frequency $\omega_c$. This frequency deviates from the
central frequency around the plasma frequency $\omega_p=$ 1THz of
the radiation without the B field.

Without the low-frequency or dc part, the two components $E_y$ and
$E_z$ show nearly the same strength and a constant phase
displacement of $\pi/2$, i.e., circular polarization. When we
reverse the B field to -x direction, the phase displacement is
changed to $-\pi/2$, i.e., the rotation of the CP radiation is also
reversed, as observed in Fig 2. Note that, in real applications, the
dc part of the radiation could be rapidly absorbed as soon as it
touches a material rather than the vacuum. The effective radiation
interacting with a sample should be a CP wave as shown in Fig. 2(a).

Now, we explain the radiation observed. The THz radiation process
\cite{Kim,THz_OE} takes place as follows: first, a net current and
plasma are formed via ionization, the current drives an
electrostatic oscillating field in the plasma, and then this field
is converted into electromagnetic radiation at the plasma
boundaries. Without an external B field, electrons released from
atoms have velocities only along the laser polarization, say the z
direction, and therefore the generated radiation is also linearly
polarized along the z direction. With the B field along the x
direction, the electrons rotate in the y-z plane and then have
velocities in both the y and z directions. Hence, the radiation has
the components along both the y and z directions.

\emph{Frequency and waveform}.$-$We set $E_y$ and $E_z$ as the
radiation or oscillation electric fields formed in plasma. The
nonrelativistic motion equation of an electron is
$dv_y/dt=-eE_y/m_e-\omega_c v_z$ and $dv_z/dt=-eE_z/m_e+\omega_c
v_y$, where both the external B field and the radiation fields much
below relativistic strengthes are considered. One easily derives
\begin{eqnarray}\label{motion}
\frac{d{\cal V}}{dt}=-\frac{e{\cal E}}{m_e}-i\omega_c {\cal V},
\end{eqnarray}
where ${\cal V}=v_y+iv_z$, ${\cal E}=E_y+iE_z$, and $i$ is the
imaginary unit. According to the wave equation
$(\nabla^2-\partial^2/c^2\partial t^2)\mathbf{E}=-(4\pi e
n_e/c^2)d\mathbf{v}/dt$ in plasma with a density $n_e$, one obtains
\begin{eqnarray}\label{wave_E}
\left(\nabla^2-\frac{1}{c^2}\frac{\partial^2}{\partial
t^2}\right){\cal E}= - \frac{4\pi e n_e}{c^2} \frac{d{\cal V}}{dt}.
\end{eqnarray}
We first use Eq. (\ref{wave_E}) omitting the spatial differential
term to look for the oscillation frequencies of the radiation source
at a given position. Set ${\cal V}$ and ${\cal E}$ with a frequency
$\omega$. According to Eqs. (\ref{motion}) and (\ref{wave_E}), one
obtains:
\begin{eqnarray}\label{frequency}
\omega_{\pm}=\sqrt{\omega_p^2+\frac{\omega_c^2}{4}} \pm
\frac{\omega_c}{2}.
\end{eqnarray}
Note that one can derive $\omega=\omega_p$ from Eqs. (\ref{motion})
and (\ref{wave_E}) in the same way for $B_0=0$. The three
frequencies $\omega_p$ and $\omega_{\pm}$ correspond to the cutoff
frequencies for wave propagation in unmagnetized and magnetized cold
plasmas, respectively \cite{Bplasma_physics}. The B field separates
the oscillation frequency from $\omega_p$ into two frequencies: one
above $\omega_p$ and the other below it. In particular, when
$\omega_c\gg \omega_p$, $\omega_+$ approaches $\omega_c$, which
provides a robust method to control the radiation frequency by the
B-field strength. The simulation results in Fig. 2 and the red line
in Fig. 3(b) are in good agreement with Eq. (\ref{frequency}). The
theoretical values are 5.19 THz and 0.19 THz, compared to 5 THz and
0 THz in the simulations with the numerical resolution of 0.5 THz.

Performing Fourier transform to Eqs. (\ref{motion}) and
(\ref{wave_E}), one obtains the dispersion relation of the radiation
wave along the -x direction, which has the refractive index
$\eta=\sqrt{1-\omega_p^2/(\omega^2+\omega\omega_c})$. Under the
condition of $\omega_+\simeq\omega_c\gg \omega_p$, $\eta\simeq1$,
and therefore, the $\omega_+$ component of the radiation generated
even in deep plasma can propagate to the vacuum with little
attenuation \cite{P.Gibbon}. Hence, the radiation is many-cycle and
narrow-band as shown in Figs. 1 and 2. This is different from the
single-cycle and broadband radiation without the B field because of
its central frequency at $\omega_p$ and $\eta=0$.

\emph{Polarization}.$-$Since high-B fields are required by
frequency-tunable radiation, we consider such B fields as
\begin{eqnarray}\label{B_condition}
\omega_L\gg \omega_c \gg \omega_p,
\end{eqnarray}
where $\omega_L$ is the laser fundamental frequency. With $\omega_c
\gg \omega_p$, the B field dominates over the plasma oscillation,
and thus, the velocity of an electron satisfies
$v_{y,j}=v_j\cos(\omega_ct+\theta_j)$ and
$v_{z,j}=v_j\sin(\omega_ct+\theta_j)$. With $\omega_L\gg \omega_c$
the initial phase $\theta_j$ can be considered as roughly the same
for all electrons, since they are released only at the laser peak
within a few cycles \cite{THz_PRE} when 50fs laser duration is used
here. Then the average electron velocity just after the passage of
the pulses can be written by
\begin{eqnarray}\label{v0}
 \left\{ { \begin{array}{l}
 {v_{y0}=v_0\cos(\omega_ct+\theta_0), }\\
 {v_{z0}=v_0\sin(\omega_ct+\theta_0). }
 \end{array}} \right.
\end{eqnarray}
We replace the electric fields with the vector potentials $A_y$ and
$A_z$ in Eqs. (\ref{motion}) and (\ref{wave_E}). From the two
equations, one obtains $d{\cal V}/dt=(e/m_ec)d{\cal A}/dt-i\omega_c
{\cal V}$ and $(\nabla^2-\partial^2/c^2\partial t^2){\cal A}=(4\pi e
n_e/c){\cal V}$, respectively, where ${\cal A}=A_y +iA_z$. Here we
are interested in the higher-frequency component with $\omega_+\gg
\omega_p$. This component with $\eta\simeq1$ can propagate in the
plasma as in a vacuum, which could be considered as a plane wave.
Therefore, the corresponding electron velocity follows $-i\omega_+
{\cal V}\simeq d{\cal V}/dt$ and $-i\omega_c {\cal V}\simeq
(\omega_c/\omega_+) d{\cal V}/dt$, with which the motion equation is
rewritten by $(1-\omega_c/\omega_+)({\cal V}-{\cal
V}_0)=(e/m_ec){\cal A}$. Inserting this equation of motion into the
wave equation expressed by ${\cal A}$, one obtains
\begin{eqnarray}\label{wave_A}
\left[\nabla^2-\frac{1}{c^2}\frac{\partial^2}{\partial
t^2}-\frac{\omega_p^2}{c^2(1-\omega_c/\omega_+)}\right]{\cal A}=
\frac{4\pi e n_e}{c} {\cal V}_0,
\end{eqnarray}
where ${\cal V}_0$ is given by Eq. (\ref{v0}). Equation
(\ref{wave_A}) describes radiation generation from a system forced
by an temporally varying external source. It is difficult to solve
analytically although a solution was given in \cite{THz_PoP} under
the condition of $B_0=0$ due to the source term independent of time.
Obviously both the y and z components of the radiation have a
strength linearly proportional to $v_0$ according to Eq. (\ref{v0}).
The two components have a phase displacement fixed at $\pi/2$ which
is determined by the one between $v_{y0}$ and $v_{z0}$. Therefore,
the radiation at $\omega_+$ is CP. The rotation of ${\cal V}_0$ and
the radiation will be reversed provided the B field sign is changed.
These agree with the simulation results above.

\begin{figure}[htbp]
\includegraphics[width=3.2in,height=2.8in]{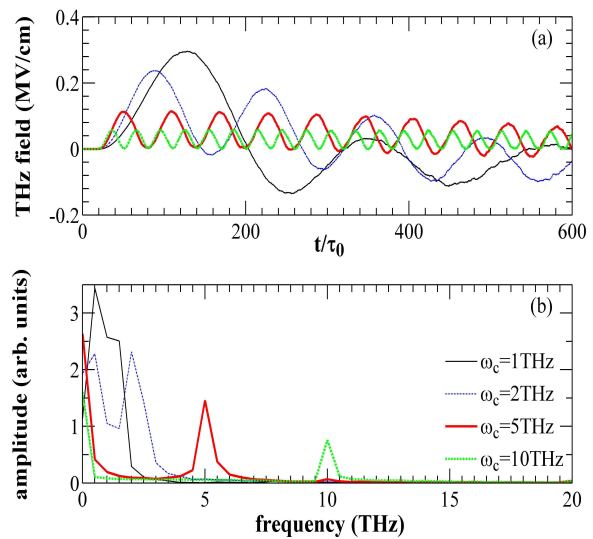}
\caption{\label{fig:epsart} (a) Temporal waveforms of the THz
electric fields $E_y$ on the axis and (b) the corresponding spectra,
where different lines correspond to different $\omega_c$
($\omega_c=$ 1THz corresponds to $B_0=$ 35.7T). }
\end{figure}

To further study the radiation features, we vary the B-field
strength and gas density in the following simulations as shown in
Figs. 3 and 4. Figure 3 illustrates that with an enhanced B field
and $\omega_c=10$ THz, the radiation frequency, polarization, and
waveform agree well with the analysis since the condition given in
Eq. (\ref{B_condition}) is sufficiently met. With $\omega_c=2$ THz
the simulation results roughly agree with the analysis. When
$\omega_c=\omega_p=1$ THz, the radiation still has the $E_y$
component and its spectrum within 0.5-1.5 THz agrees with Eq.
(\ref{frequency}). Its waveform attenuates with time, approaching
the one without the B field, because its frequency is close to
$\omega_p$.

\begin{figure}[htbp]
\includegraphics[width=3.2in,height=2.8in]{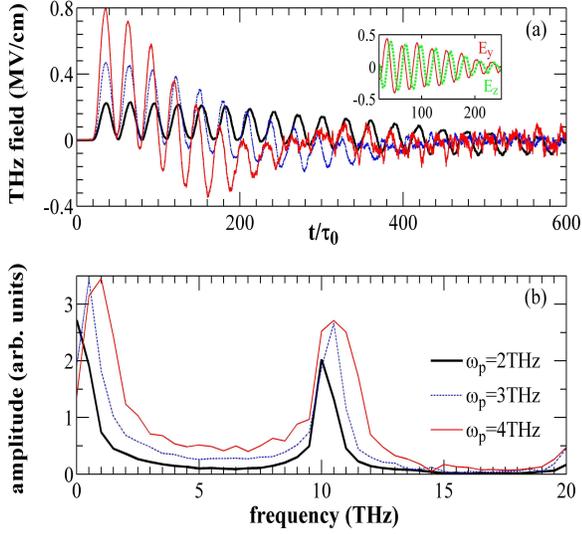}
\caption{\label{fig:epsart}(a) Temporal waveforms of the THz
electric fields $E_y$ on the axis and (b) the corresponding spectra,
where different lines correspond to different $\omega_p$ or gas
densities. The B field strength is fixed at 357T. The inset in (a)
shows the radiation fields $E_y$ and $E_z$ with $\omega_p=$ 4 THz
and the components below 2 THz are filtered.}
\end{figure}

In Fig. 4 we take different gas densities: $4.9\times 10^{16}~\rm
cm^{-3}$, $1.1\times 10^{17}~\rm cm^{-3}$, and $1.96\times
10^{17}~\rm cm^{-3}$ with the corresponding $\omega_p=$ 2, 3 and 4
THz. The radiation frequencies agree well with Eq.
(\ref{frequency}). The radiation has a many-cycle waveform but the
temporal attenuation of the waveform becomes more obvious with the
growing density since the frequency $\omega_+$ is closer to
$\omega_p$. The radiation is nearly CP even with $\omega_p=$ 4 THz,
as observed in the inset in Fig. 4(a), provided the lower-frequency
component is filtered. Note that as $\omega_-$ increases with the
growing $\omega_p$, the field envelope tends to oscillate around the
axis ($E_y=0$).

\begin{table}[h]
\caption{Strength (MV/cm) of CP THz radiation of the $\omega_+$
component as a function of $\omega_c$ (the row) and $\omega_p$ (the
column), where $\omega_c=$ 2THz corresponds to $B_0=$ 71T. }
\begin{tabular}{l l l l l l l l}
 \hline\hline
 ~~~~~~&  ~~2 THz& ~~5 THz&  ~~10 THz&  ~~15 THz&  ~~20 THz& ~~30 THz&\\
 \hline
 1 THz&   ~~0.13& ~~0.054&  ~~0.027&  ~~0.016& ~~0.013&    ~~0.009&\\
 2 THz&   ~~&     ~~&       ~~0.12&\\
 3 THz&   ~~&     ~~&       ~~0.23&\\
 4 THz&   ~~&     ~~&       ~~0.41&    ~~0.27&   ~~0.20&   ~~0.12&\\
 \hline\hline
\end{tabular}
\end{table}

We list the radiation strengths obtained in simulations as a
function of $\omega_c$ and $\omega_p$ in Table I. It is shown that
the radiation strength roughly follows:
\begin{eqnarray}\label{scaling}
E_{THz}^{\omega_+} \propto \frac{\omega_p^2}{\omega_c}.
\end{eqnarray}
According to Eq. (\ref{wave_A}), the strength scales linearly with
the plasma density or the net current strength, i.e.,
$E_{THz}^{\omega_+} \propto \omega_p^2$. With a given plasma
density, the current strength is nearly not changed with the B
field. Multiplying the electron motion equation by the electron
velocity $\mathbf{v}$, one obtains
$dv^2/dt=-2e\mathbf{E}\cdot\mathbf{v}/m_e$, where $\mathbf{E}$ is
the laser electric fields. When the B field satisfies Eq.
(\ref{B_condition}) and tens of fs laser durations are considered
here, the rotation of $\mathbf{v}$ from the laser polarization plane
is slight during the laser interaction with the electron and
therefore, the net gain of the electron energy (also the radiation
energy) is nearly the same as the case without the B field. With a
given radiation energy, the radiation strength will decrease
linearly with its frequency $\omega_+ \simeq\omega_c$, i.e.,
$E_{THz}^{\omega_+} \propto 1/\omega_c$.

\begin{figure}[htbp]
\includegraphics[width=3.2in,height=2.8in]{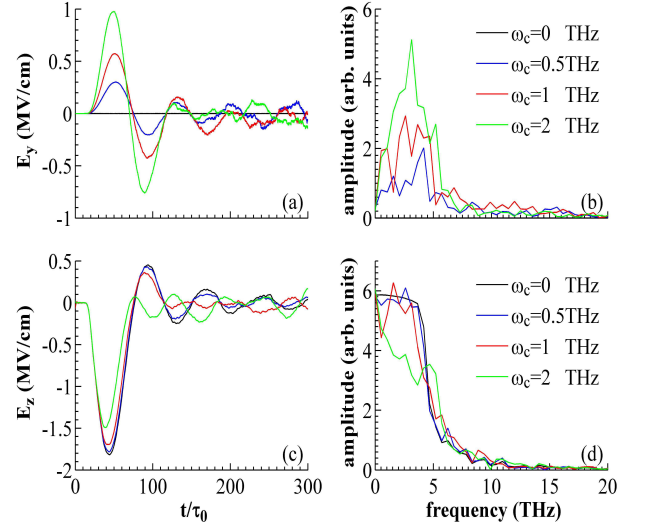}
\caption{\label{fig:epsart} [(a) and (c) ]Temporal waveforms of the
THz electric fields on the axis and [(b) and (d)] the corresponding
spectra, where the gas density is taken as $1.96\times 10^{17}~\rm
cm^{-3}$ ($\omega_p=$ 4THz) and different lines correspond to
different $\omega_c$ ($\omega_c=$ 0.5THz corresponds to $B_0=$
17.8T). }
\end{figure}

Next, we consider relatively low B fields with $\omega_c<\omega_p$
as shown in Fig. 5. EP radiation is also generated even with $B_0=$
17.8T or $\omega_c=$ 0.5THz. The amplitude of $E_y$ grows with the
B-field strength since more electron energy is transferred to the y
direction by the B field. Both $E_y$ and $E_z$ have near
single-cycle waveforms since their frequencies are close to
$\omega_p$. The two-frequency spectrum disappears due to a low value
of $\omega_c$. The relative phase is difficult to calculate since
the radiation is broadband. We filter the low-frequency part below
1THz (such filter could be more meaningful for a spectrum with two
frequencies separated) and the relative phase is changed from
$0.8\pi$ to around $0.5\pi$ as $\omega_c$ grows from 0.5THz to 4THz.

The radiation with either a high or low B field is generated due to
the gyrational motion of plasma electrons under the B field. This
field slightly affects the gas ionization responsible for the
current formation. Hence the magnetic approach can be extended to
other laser-plasma-based THz emission schemes, which has been
verified by our simulations with an asymmetric-laser scheme
\cite{THz_PoP}. Besides, we have performed a 3D PIC simulation and
observed the same result as a 2D simulation with the same parameters
as in Fig. 1 but with a plane laser driver. The result also
approaches that with $r_0=150\rm \mu m$ in Fig. 1(c), indicating
that our study is valid with a larger spot radius $r_0$.

In summary, we have demonstrated a unique EP or CP, narrow-band THz
source if a static B field is applied. With a high B field at a 100T
scale, the radiation shows two frequencies: the lower is nearly dc
and the higher (central frequency) almost at $\omega_c$. Therefore,
the central frequency can be adjusted linearly by the B-field
strength. The radiation rotation can also be controlled by the
B-field sign. With the B field decreased to the 10T scale, EP
radiation is still generated but becomes broadband and single cycle.
To fully apply this scheme under the high-B-field condition in Eq.
(4) to the whole THz band, the B-field strength should be tunable
within 3.57T to 357T (corresponding to 0.1THz to 10THz). Besides the
B-fields generated in traditional ways
\cite{Labs_highB,B600T,B310T}, one may employ novel
nanosecond-laser-driven superstrong B-fields
\cite{Fujioka,Lasest_Bfield} with the strength continuously tunable
by the laser energy and with a duration of nanoseconds (far longer
than THz radiation cycle). With this source, one can realize the
magnetically controlled two-color scheme in an all-optical way,
where the B-field generation is easily synchronized with the
two-color laser driver.

\begin{acknowledgments}
W. M. W. acknowledges support from the Alexander von Humboldt
Foundation. The authors gratefully acknowledge the computing time
granted by the JARA-HPC and VSR committees on the supercomputers
JUROPA and JUQUEEN at Forschungszentrum J\"ulich. This work was
supported by the National Basic Research Program of China (Grants
No. 2013CBA01500 and 2014CB339800) and NSFC (Grants No. 11375261,
11421064, 11411130174, 11375262 and 11135012).
\end{acknowledgments}

%
%\newpage
%\centerline{\bf Figure Captions} \vskip 0.5cm
%
%
%Fig. 1. [(a) and (b)] Snapshots of the THz electric fields (MV/cm)
%at the time of 0.7 ps, and (c) the field distributions on the axis
%($y=0$) at 2 ps, where an external B field of 178T is imposed. The
%broken lines in (c) correspond to the case
%without the B field.\\
%
%
%Fig. 2. (a) Temporal waveforms of the THz electric fields on the
%axis and (b) the corresponding spectra, where an external B field of
%178T is imposed. The components below 1 THz
%are filtered.\\
%
%
%Fig. 3. (a) Temporal waveforms of the THz electric fields $E_y$ on
%the axis and (b) the corresponding spectra, where different lines
%correspond to different $\omega_c$
%($\omega_c=$ 1THz corresponds to $B_0=$ 35.7T).\\
%
%
%Fig. 4. (a) Temporal waveforms of the THz electric fields $E_y$ on
%the axis and (b) the corresponding spectra, where different lines
%correspond to different $\omega_p$ or gas densities. The B field
%strength is fixed at 357T. The inset in (a) shows the radiation
%fields $E_y$ and $E_z$ with $\omega_p=$ 4 THz
%and the components below 2 THz are filtered.\\
%
%
%Fig. 5. [(a) and (c)] Temporal waveforms of the THz electric fields
%on the axis and [(b) and (d)] the corresponding spectra, where the
%gas density is taken as $1.96\times 10^{17}~\rm cm^{-3}$
%($\omega_p=$ 4THz) and different lines correspond to different
%$\omega_c$ ($\omega_c=$ 0.5THz corresponds to $B_0=$ 17.8T).

\end{document}